\documentclass[twocolumn,amsmath,amssymb,floatfix]{revtex4}
\usepackage{graphicx}
\usepackage{bm}

\def\Vec#1{\boldmath {#1}}
\def\hsigma{\hat{\sigma}}
\def\hs{\hat{s}}
\def\TR{\mathrm {Tr}} 
\def\sgn{\mathrm{sgn}}

\begin{document}
\sloppy

\title{Naive mean field approximation for image restoration}
\author{Hayaru Shouno}
\email{shouno@ics.nara-wu.ac.jp}
\affiliation{Graduate School of Human Culture, Nara Women's University}

\author{Koji Wada}
\affiliation{Graduate School of Science and Engineering, Saitama University}

\author{Masato Okada}
\affiliation{Brain Research Institute, RIKEN}

\date{\today}

\begin{abstract}
We attempt image restoration in the framework of the Bayesian inference.
Recently, it has been shown that under a certain criterion
the MAP (Maximum A Posterior) estimate,
which corresponds to the minimization of energy,
can be outperformed  
by the MPM (Maximizer of the Posterior Marginals) estimate,
which is equivalent to a finite-temperature decoding method.
Since a lot of computational time is needed for the MPM estimate
to calculate the thermal averages,
the mean field method, which is a deterministic algorithm,
is often utilized to avoid this difficulty.
We present a statistical-mechanical analysis of
naive mean field approximation
in the framework of image restoration.
We compare our theoretical results with those of computer simulation,
and investigate the potential of naive mean field approximation.
\end{abstract}

\maketitle                        

\section{Introduction}
In this paper, we investigate the image restoration problem.
This problem involves
synthesis of an image from a corrupted image with a model of the
information available (or assumed) on the source image and
the corruption process \cite{Geman84} \cite{Pryce95} \cite{Geiger91}.
%
%
%
The problem lends itself naturally to a Bayesian formulation, which 
estimates a probability (posterior) for the original image 
on the basis of the model probability (priors) of the assumed model of 
the source image and corruption process.

%
One strategy in image restoration is to use the Bayesian inference by 
adopting the image that maximizes the posterior probability.
This method is called the maximum a posteriori (posterior) probability
(MAP) inference.
Given a corrupted input image, the MAP inference accepts the image
that maximizes the posterior probability as the restored result.
The logarithm of posterior probability can be regarded as the energy, 
so we can consider the MAP inference as an energy minimization problem.
Therefore, the image restoration problem can be regarded as an
optimization problem.
Geman \& Geman demonstrated that the MAP inference result can be applied to 
the image restoration problem by using simulated annealing, 
which is a tool for searching the ground state \cite{Geman84}.

Another strategy is the inference in which the expectation value with
respect to the maximized marginal posterior probability at each site in
thermal equilibrium is regarded as the original image. 
This method is called maximizer of the posterior marginals (MPM)
inference \cite{Marroquin87} \cite{Rujan93} \cite{Sourlas94}. 
In the MAP inference, the posterior probability is given for 
each set of pixels.
In contrast, in the MPM inference, 
the posterior marginal probability controlled by the temperature $T$ 
is given for each pixel value.
To find a restoration image by the MPM inference, 
we should calculate thermal average for each pixel value. 
The MPM inference includes the MAP inference \cite{Nishimori00} \cite{NishimoriBook01}
because
, at the limit of the temperature $T\rightarrow 0$, 
the MPM inference becomes equivalent to the MAP inference.
In general, however, the MPM inference is evaluated at $T>0$, 
so this method is also called ``finite temperature restoration'',
and the MPM inference has an advantage over the MAP inference 
in the restoration ability of each pixel \cite{Nishimori00} \cite{NishimoriBook01}.
\begin{figure*}   
\begin{center}
  \resizebox{0.95\textwidth}{!}
  {\includegraphics{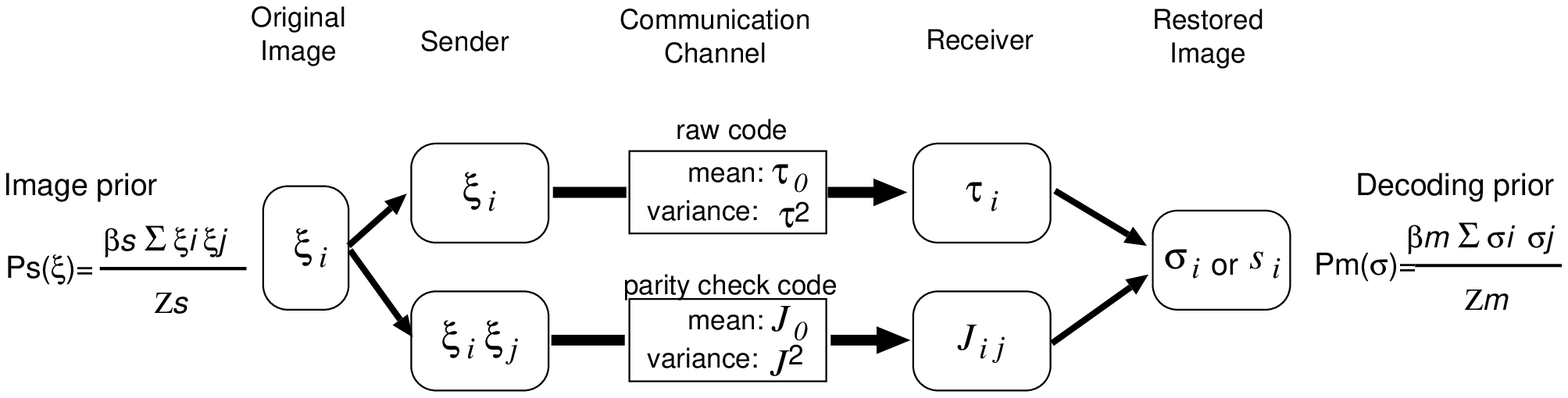}}
 \caption{Schematic diagram of image restoration problem}
 \label{fig:method}
\end{center}
\end{figure*}

%
%
Recently, the MPM inference has been discussed in the field of error
correcting code \cite{Rujan93} \cite{Sourlas94} \cite{Nishimori00}.
The problem of error correcting code is similar to the image restoration
problem in the sense that
the received bit sequence, 
which corresponds to the image represented by a set of pixels,
is corrupted by noise, 
and
the receiver tries to retrieve the original bit sequence/image
from the noisy one.
The major difference between error correcting code 
and image restoration is that
the image restoration problem usually gives only the corrupted
image and not other additional redundant information.
Therefore, in image restoration, we usually assume 
alternative information for the original image such as 
the prior probability.
In the field of error correcting code, 
Rujan has pointed out that 
carrying out the decoding procedure not at the ground state 
but at a finite temperature is effective \cite{Rujan93}.
Sourlas used the Bayes formula to re-derive the finite-temperature
decoding of Rujan's result under more general conditions \cite{Sourlas94}.
%
%

Assuming the number of pixels as $N$ and each pixel as a binary unit 
that can take $\{-1,+1\}$, 
the number of feasible combinations of pixel values becomes $2^N$.
Because of the existence of local minima, finding the ground state task 
is very difficult with a method such as the gradient decent.
Thus, to apply the MAP or MPM inference, 
we usually use a relaxation method as typified by annealing technique.
%
%
Moreover, the MPM inference involves a more important problem 
as described below.
In the MAP inference, once the system reaches 
the macroscopic equilibrium state,
each pixel value can be properly determined with its probability as $1$.
On the other hand, in the MPM inference, when the system reaches
the macroscopic equilibrium state, 
each pixel value cannot be determined uniquely
because the probabilities of each binary state have some finite values 
in the finite temperature decoding.
Therefore, we should calculate thermal averages for each pixel, 
and this requires many samplings.

In this study, we discuss an approximation that replaces
the stochastic dynamics of the MPM inference with 
deterministic dynamics.
In statistical mechanics, this approximation
is called the ``naive mean field equation (NMFE)'' \cite{Bray86}.
The NMFE has usually been applied for
emulating the behavior of the stochastic unit.
By Applying the NMFE, each restoring unit, 
which is defined as a stochastic binary unit,
is replaced by a deterministic analog unit that takes $[-1,+1]$.
One important advantage of applying NMFE is the ability to
reduce calculation cost 
by eliminating the need to calculate thermal average
, which requires many samplings.
The NMFE has been applied to several combinatorial optimization problems
such as the traveling salesman problem (TSP) \cite{Hopfield86}.
%
%
However, almost all researchers who have used NMFE 
have not pursued quantitative issues such as the accuracy of
the approximation.

In \S \ref{sec:model}, we formulate the image restoration problem using the 
Bayes inference in the manner of Nishimori \& Wong's
formulation\cite{Nishimori00}. 
Recently, from the statistical-mechanical point of view, 
Nishimori \& Wong have formulated the image restoration problem
by introducing a mean field model for binary image restoration.
They analyzed the model theoretically by the replica method.
We applied NMFE to the formulation and analyzed it by the replica method in
the manner of Bray {\it et al.} \cite{Bray86}.

In \S \ref{sec:simulation}, we compare the results between 
our analysis and computer simulations.
Within the limits of the mean field approximation, 
our analysis showed agreement with the computer simulations.
However, the prior probability derived from the mean field
approximation, which is called an infinite range model,
is usually not practical for a real image prior.
Therefore, we discuss the difference between infinite range model and 
its origin, called the nearest 0 interaction model.

\section{Model and Analysis}
\label{sec:model}
\subsection{Formulation of the Image restoration}

In this section, we apply the mean field theory to 
the image restoration problem in the manner of Nishimori \& Wong.
Figure \ref{fig:method} shows a schematic diagram of image restoration.
The original image is represented by $\{\xi_i\}$, 
and each unit is a binary unit that takes two states $\{-1, +1\}$.
The number of units corresponding to the image size is denoted by $N$.
In addition, the original image $\{\xi_i\}$ is assumed to have the
following prior probabilities:
\begin{align}
 P_s \left(\left\{ \xi_i \right\}\right) &=
  \frac{1}{Z(\beta_s)}
  \exp( \frac{\beta_s}{N} \sum_{i<j} \xi_i \xi_j),
  \label{eq:orig_prior}
  \\
 Z(\beta_s) &= \TR_{\Vec{\xi}}
  \exp( \frac{\beta_s}{N} \sum_{i<j} \xi_i \xi_j),
\end{align}
where $\beta_s > 0$.
The operator $\TR_{\Vec{\xi}}$ means trace, 
a sum over all possible $2^N$ states of $\Vec{\xi}$.
$Z(\beta_s)$ indicates the partition function of this prior probability.
Assuming $\beta_s >0$, this probability suggests that pixel values $\xi_i$
and $\xi_j$ have a tendency to take the same value.

In image restoration, the interaction between each pixel 
should be described by a local rule, such as a sum of the nearest neighbors.
For analysis by the mean field theory, however, 
we replace this local interaction rule with a global one, 
that means each pixel has interactions to all other pixels.

In the conventional image restoration framework, 
the transmitting signal is only the raw code $\Vec{\xi}$ (upper path in 
Fig.\ref{fig:method}). 
However, considering the similarity of the formulation of the 
error-correcting code problem using a Bayesian inference
\cite{Rujan93} \cite{Sourlas94} \cite{Nishimori00},
it is natural to introduce transmission of  a redundant code 
for better image restoration.
Nishimori \& Wong proposed the introduction of a redundant code 
called the ``parity check code'' (lower path in Fig. \ref{fig:method})
as well as the raw code.  
For comparison with their results, in this study 
we also adopted transmission of the parity check code.
For the parity check code, we adopted a 2-body interaction term 
denoted by $\xi_i\xi_j / N$.
The reason for dividing $\xi_i\xi_j$ by $N$ is to apply  the 
mean field theory. 
At the receiver side, 
the transmitting signal $\xi_i$ corresponds to $\tau_i$, 
and 
the parity check code $\xi_i\xi_j/N$ corresponds to $J_{ij}$.
The posterior probability based on observation signal
$P(\{\xi_i\} | \{J_{ij}\}, \{\tau_i\})$ can be represented by
the Bayes formula:
\begin{equation}
 P(\{\xi_i\} | \{J_{ij}\}, \{\tau_i\}) \propto 
  P_{\mathrm{out}}( \{J_{ij}\}, \{\tau_i\} | \{\xi_i\})  
  P_s(\{\xi_i\}).
\end{equation}

When the receiver can guess the form of the original image 
posterior probability, 
the restoring posterior probability can be assumed by replacing 
the original pixel value $\xi_i$ with the
estimated pixel value $\sigma_i$.
In general, however, the receiver could not guess the original image prior
probability accurately.
In order to evaluate the performance of the restoration ability 
by the difference between the original image prior and the restoring prior,
%
we assume that the  restoring image prior probability $P_m(\cdot)$ has a
different parameter $\beta_m$ from the original image prior $P_s(\cdot)$.
\begin{align}
 P_m \left(\left\{ \sigma_i \right\}\right) &=
  \frac{1}{Z(\beta_m)}
  \exp( \frac{\beta_m}{N} \sum_{i<j} \sigma_i \sigma_j),
  \label{eq:restoreprior}
  \\
 Z(\beta_m) &= \TR_{\Vec{\sigma}}
  \exp( \frac{\beta_m}{N} \sum_{i<j} \sigma_i \sigma_j),
\end{align}
where $\beta_m > 0$.
Each form of prior is identical mathematically
except for the parameters of interaction strength.
We primarily discuss the $\beta_s \neq \beta_m$ case.

$P_{\mathrm{out}}( \{ J_{ij} \}, \{\tau_i\} | \{\xi_i\})$
is a conditional probability of the observation signal, and
it is a probability expression of the corrupting process in the 
transmission channel. 
In this study, we assume that the noise added in the corrupting process 
is similar to Gauss distribution as follows:  
\begin{align}
 & P_{\mathrm{out}}( \{J_{ij}\},\{\tau_i\}|\{\xi_i\})  
 \propto 
 \notag\\
 &
 \quad
 \exp
  \biggl(
   -\frac{N \sum
   \left( J_{ij} - \frac{J_{0}}{N} \xi_i \xi_j \right)^2 
   }{2J^2}
  -\frac{ \sum
  \left( \tau_i - \tau_0 \xi_i \right)^2
  }
  {2\tau^2} 
 \biggr)
 \label{eq:noise}
\end{align}
The first term in the exponential indicates that the noise corresponds to
the parity check code, and the second term corresponds to the raw code.
For the first term in the exponential, the random variables $J_{ij}$
follow the normal distribution, whose average is $J_0 \xi_i \xi_j /N$,
and the variance is $J^2/N$.
In the second term in the exponential, the random variables $\tau_i$ 
follow the normal distribution whose average is $\tau_0\xi_i$, and the
variance is $\tau^2$.

Nishimori \& Wong discussed the macroscopic characteristic of this system
by using statistical mechanics \cite{Nishimori00}. 
They introduced a system that has the following Hamiltonian:
\begin{equation}
 H = - \beta \sum_{i<j} J_{ij} \sigma_i \sigma_j 
  - \frac{\beta_m}{N} \sum_{i<j} \sigma_i \sigma_j 
  -h \sum_{i} \tau_i \sigma_i.
  \label{eq:hamiltonian}
\end{equation}
They calculated the free energy by the replica method for averaging
probabilities
$P_s \left(\left\{ \xi_i \right\}\right)$  and
$P_{\mathrm{out}}( \{J_{ij}\}, \{\tau_i\} | \{\xi_i\})$.
Ignoring the 2-body interaction term, that is $\beta=0$,
this Hamiltonian consists of 
the prior probability term 
and 
the observed data fitting term.
This form is equivalent to the conventional cost function form.

\begin{figure}[t]   
\begin{center}
 \resizebox{4cm}{!}
 {\includegraphics{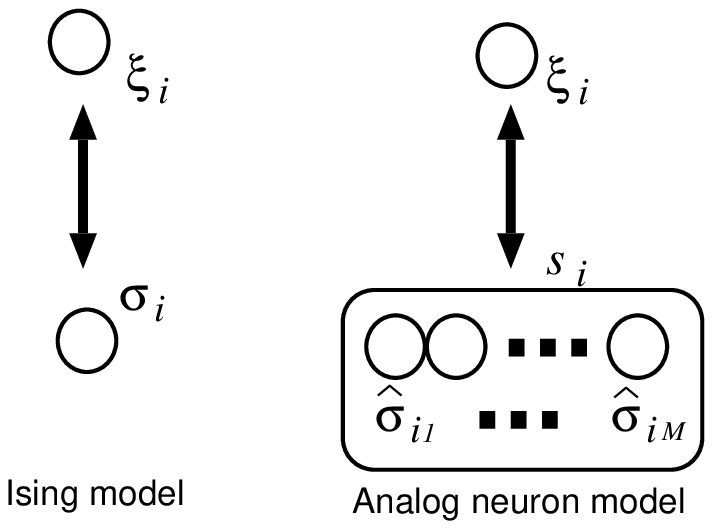}}
 \caption{Analog neuron decoder for analysis:
 In Ising model, encoding unit $\xi_i$ corresponds to decoding unit $\sigma_i$.
 On the contrary, the analog model assumes decoding unit have $M$ units, and
 their average is regarded as the analog unit output.
 }
 \label{fig:analog}
\end{center}
\end{figure}

In the MAP inference, 
the estimation value of restored pixels $\Vec{\sigma}$ results in 
the minimum of the Hamiltonian 
defined in eq. (\ref{eq:hamiltonian}).
On the other hand, the MPM inference, 
which corresponds to the finite temperature estimation,
estimates pixel values by the 
$\sgn(\langle \sigma_i \rangle)$,
where $\langle \sigma_i \rangle$ means
the thermal average.
Considering the limit of the temperature at $0$, the MPM inference 
becomes equivalent to the MAP inference.

We introduced the quantity overlap $M_o$ to evaluate 
the restoration ability measure for the MPM inference.
\begin{equation}
 M_o = \frac{1}{N} \sum_i \xi_i \sgn(\langle\sigma_i\rangle).
  \label{eq:overlap}
\end{equation}
Nishimori \& Wong suggested that
the minimization of the Hamiltonian (\ref{eq:hamiltonian}) and 
the maximization of the overlap 
are not equivalent \cite{Nishimori00}.
Moreover they also indicated that 
the MPM inference (finite temperature estimation)
has better restoration ability 
than the MAP inference in the sense of the overlap:
\begin{equation}
 \begin{split}
  M_o &= \TR_{\Vec{\xi},\Vec{J},\Vec{\tau}}
  P_s \left(\left\{ \xi_i \right\}\right)
  P_{\mathrm{out}}( \{J_{ij}\}, \{\tau_i\} | \{\xi_i\})  
  \\
  & \quad\quad\quad\quad 
  \times \xi_i \sgn(\langle \sigma_i \rangle).
 \end{split}
\end{equation}
This overlap can be evaluated by averaging probability variables with 
the statistical-mechanical technique.

\subsection{Naive mean field approximation}
%
%
In the finite temperature decoding which corresponds to the MPM inference, 
we should calculate the thermal average $\langle \sigma_i \rangle$,
which means the estimation value of $\sigma_i$.
We evaluated the calculation cost of MPM inference by computer
simulation.
As we will show that calculating this average requires
about 50 times longer computational time than that
for achieving to the macroscopic equilibrium state.
This result will be given in \S \ref{sec:simulation}.


In this study, to find the ground state,
we introduce the analog neural network model for approximation 
proposed by Hopfield \& Tank \cite{Hopfield86}.
By this approximation, 
each Ising unit is replaced by an analog unit 
that can take continuous value $[-1,+1]$,
and the output of each analog unit can be regarded as the thermal
average of the corresponding Ising spin unit,
which can take two states $\sigma_i = \pm 1$.
%
%
For replacing Ising unit to analog one, 
we introduced a Hamiltonian as substitution for 
eq. (\ref{eq:hamiltonian})
:
\begin{equation}
 {\cal H} = M\left[ - \beta  \sum_{i<j} J_{ij} \hs_i \hs_j 
  - \frac{\beta_m }{N} \sum_{i<j} \hs_i \hs_j 
  -h \sum_{i} \tau_i \hs_i \right],
  \label{eq:Bray_hamiltonian}
\end{equation}
where the unit $\hs_i$ is an analog unit.
$M$ is a scaling factor described as below.
Following the manner of Bray {\it et al}\cite{Bray86}, 
we assumed each $i$th site consists of $M$ Ising units, and 
the analog unit output $\hs_i$ is calculated by 
the average of $M$ Ising units $\hsigma_{ia}$ (see Fig. \ref{fig:analog}).
\begin{equation}
 \hs_i = \frac{1}{M} \sum_{a}^{M} \hsigma_{ia}.
  \label{eq:heikin}
\end{equation}
Considering the limit  $M\rightarrow\infty$, each output $\hs_i$ can
take an analog value $[-1,+1]$.
When $M$ is a finite value,
the analog unit output $\hs_i$ is called 'binominal spins' which 
can take $-1, -1+\frac{2}{M}, \cdots, 1-\frac{2}{M}, 1$ with
binominal distribution.
So that, we can define `spin weight function' as
\begin{align}
 W(\hs_i) &= M \:\: \TR \:\: \delta \left( M\hs_i - \sum_{a=1}^{M} \hsigma_{ia}
 \right) 
 \notag\\
 &= \frac{M}{2\pi {\mathrm j}} 
 \int_{-{\mathrm j} \infty}^{+{\mathrm j}\infty} du_i 
 \exp( M (-u_i \hs_i + \ln 2 \cosh(u_i) )).
\end{align}
The partition function $Z$ can be described as
\begin{align}
 Z &= \prod_{i=1}^{N} \int_{-1}^{+1} d\hs_i W(\hs_i) \exp( {\cal H} ) 
 \notag\\
 &=
 M  \int_{-{\mathrm j} \infty}^{+{\mathrm j}\infty} 
 \prod_{i=1}^{N} \left(\frac{du_i}{2\pi j}\right)
 \int_{-1}^{+1} \prod_{i=1} d\hs_i
 \notag\\
 & \quad
 \exp( M (\beta \sum_{i<j} J_{ij} \hs_i \hs_j + \frac{\beta_m}{N} \hs_i
 \hs_j + h \sum_i \tau_i \hs_i
 \notag\\
 & \quad\quad
 -u_i \hs_i + \ln 2 \cosh(u_i) )) 
 \label{eq:analogZ}
\end{align}
In the limit $M\rightarrow \infty$, the integrals over $\{u_i\}$ and
$\{\hs_i\}$ in eq. (\ref{eq:analogZ}) could be evaluated by the
saddle-point method.
The stationary equations are
\begin{align}
 0 &= \beta \sum_{i<j} J_{ij} \hs_i \hs_j + \frac{\beta_m}{N} \sum_{i<j}
 \hs_i \hs_j + h \sum_i \tau_i \hs_i - u_i,
 \\
 0 &= -\hs_i + \tanh u_i.
\end{align}
Eliminating $u_i$, we can obtain 
\begin{equation}
 \hs_i = \tanh (  \beta  \sum_j J_{ij} \hs_j 
  + \frac{\beta_m}{N} \sum_j \hs_j + h \tau_i ).
  \label{eq:eq}
\end{equation}

From the manner of Hopfield \& Tank \cite{Hopfield86}, 
we can also consider a discrete synchronous updating rule:
\begin{equation}
 s^{t+1}_i = \tanh (  \beta  \sum_j J_{ij} s^t_j 
  + \frac{\beta_m}{N} \sum_j s^t_j + h \tau_i ),
  \label{eq:dynamics}
\end{equation}
where $s^{t}_i$ denotes the analog unit output at time $t$.
When the system described by eq. (\ref{eq:dynamics}) 
reached to the equilibrium state $s^{\infty}_i$, 
whole units should satisfy eq. (\ref{eq:eq}).
Therefore, to investigate the equilibrium state of dynamics
(\ref{eq:dynamics}), 
we should take the analog Hamiltonian described by eq.
(\ref{eq:Bray_hamiltonian}) correctly.
In the equilibrium state, each analog unit state expressed by
$s^{\infty}_i$ 
corresponds to $\langle \sigma_i \rangle$, i.e.
$s^{\infty}_i$ can be regarded as the thermal average of $\sigma_i$.

This analog unit replacement is sometimes called 
the ``naive mean field equation (NMFE) approximation''.
From eq. (\ref{eq:dynamics}), this system follows 
deterministic dynamics, 
so that it is no need to calculate the thermal average of stochastic unit; 
therefore, the calculation cost is lower than the stochastic Ising model.

However this replacement from stochastic Ising unit $\sigma_i$ 
to the deterministic analog unit $s_i$ 
is an merely substitution.
Thus, we should analytically investigate the result of 
applying NMFE by analog model 
to compare it with the estimated thermal average of the Ising model.
Hence, to investigate the equilibrium state of the system described by 
eq. (\ref{eq:eq}), 
we should analyze the system that has the Hamiltonian described by
eq. (\ref{eq:Bray_hamiltonian}).

\subsubsection{Replica method}
We now analyze the system that has the Hamiltonian described by
eq. (\ref{eq:Bray_hamiltonian}) by the ``replica method'',
which is a standard statistical analysis tool.
The $n$ replicated partition function can be expressed as:
\begin{equation}
\begin{split}
 [Z^n] &= 
  \mathop{\TR} \left[ \int \prod_{i<j} dJ_{ij} \sqrt{\frac{N}{2\pi J^2}}
   \right]
  \left[ \int \prod_{i} \frac{d\tau_i}{\sqrt{2\pi}\tau} \right]
 \\
 &
 \times
 P_s \left(\left\{ \xi_i \right\}\right)
 P_{\mathrm{out}}( \{ J_{ij} \}, \{ \tau_i \} | \{ \xi_i \})  
 \\
 &
 \times
 \exp\biggr(
  \frac{\beta}{M} \sum_{i<j, \alpha, a, b}
  J_{ij} \hsigma_{ia}^{\alpha} \hsigma_{jb}^{\alpha}
  \\
 &
 +
  \frac{\beta_m}{MN} \sum_{i<j, \alpha, a, b}
  \hsigma_{ia}^{\alpha} \hsigma_{jb}^{\alpha}
  +\sum_{i, \alpha, a} h \tau_i \hsigma_{ia}^{\alpha}
  \biggr), 
\end{split}
\end{equation}
where operator $\TR$ means the sum over all states about
$\{\hsigma_{ia}^{\alpha}\}$ and  $\{\xi_i\}$.
We analyzed this replicated partition function by the standard replica
method. 
The replica symmetry solution can be described as:
\begin{align}
 m_0 &= \tanh( \beta_s m_0 ),
 \label{eq:m0} \\
 m   &= \frac
  {
 \mathop{\TR}_{\xi} \;  e^{\beta_s m_0 \xi}  \int Dx \; \hat{F}(U(x))
 }{2 \cosh(\beta_s m_0)},
 \\
 t   &= \frac{
  \mathop{\TR}_{\xi} \;  e^{\beta_s m_0 \xi} \xi  \int Dx \; \hat{F}(U(x))
  }{2 \cosh(\beta_s m_0)},
  \\
 q &= \frac{
  \mathop{\TR}_{\xi} \;  e^{\beta_s m_0 \xi}   \int Dx \; \hat{F}(U(x))^2
  }{2 \cosh(\beta_s m_0)},
  \\
 \chi &= 
  \frac{1}{\sqrt{h^2 \tau^2 + \beta^2 J^2 q}}
 \frac{
  \mathop{\TR}_{\xi} \;  e^{\beta_s m_0 \xi}  \int Dx \; x \hat{F}(U(x))
  }{2 \cosh(\beta_s m_0)}, \label{eq:chi}
\end{align}
where $U(\cdot)$ means
\begin{equation}
 U(x) = \sqrt{h^2 \tau^2 + \beta^2 J^2 q} x + \beta_m m + (h \tau_0 +
  \beta J_0 t )\xi,
\end{equation}
and the function $\hat{F}(\cdot)$ is a solution of the self consistent
equation: 
\begin{eqnarray}
 F(x) &=& \tanh( U(x) + \beta^2 J^2 \chi F(x) ).
  \label{eq:onsager}
\end{eqnarray}
Using these solutions, we could obtain the overlap $M_o$:
 \begin{equation}
  \begin{split}
   M_o &= \frac{1}{N} \sum_i \xi_i \mbox{sgn}(s_i) \\
   &= \frac{
   \mathop{\TR}_{\xi} \;  e^{\beta_s m_0 \xi}\int Dx \; \xi\sgn(U(x))
   }{2 \cosh(\beta_s m_0)}.
   \label{eq:Bray_overlap}
  \end{split}
 \end{equation}

\section{Results}
\label{sec:simulation}
In this section, 
we compare the analysis results between 
applying the NMFE approximation (analog model) 
and estimated thermal average of the Ising model.
We also compare these theoretical analyses
with the computer simulation results.

In the following subsection from \ref{sec:subsim1} to \ref{sec:subsim3}, 
we treated infinite range Ising/analog model.
In the infinite range model, 
the original image prior probability is defined by eq. (\ref{eq:orig_prior}), 
and 
the image is corrupted by the noisy channel according to eq. (\ref{eq:noise}).
In the infinite range Ising model,
we assumed restoring prior probability as eq. (\ref{eq:restoreprior}), 
and the Hamiltonian could be derived as eq. (\ref{eq:hamiltonian}).
On the other hand, in the infinite range analog model, we defined the
Hamiltonian as eq. (\ref{eq:Bray_hamiltonian}).

\begin{figure}[t]   
\begin{center}
 \resizebox{!}{5.5cm}
 {\includegraphics{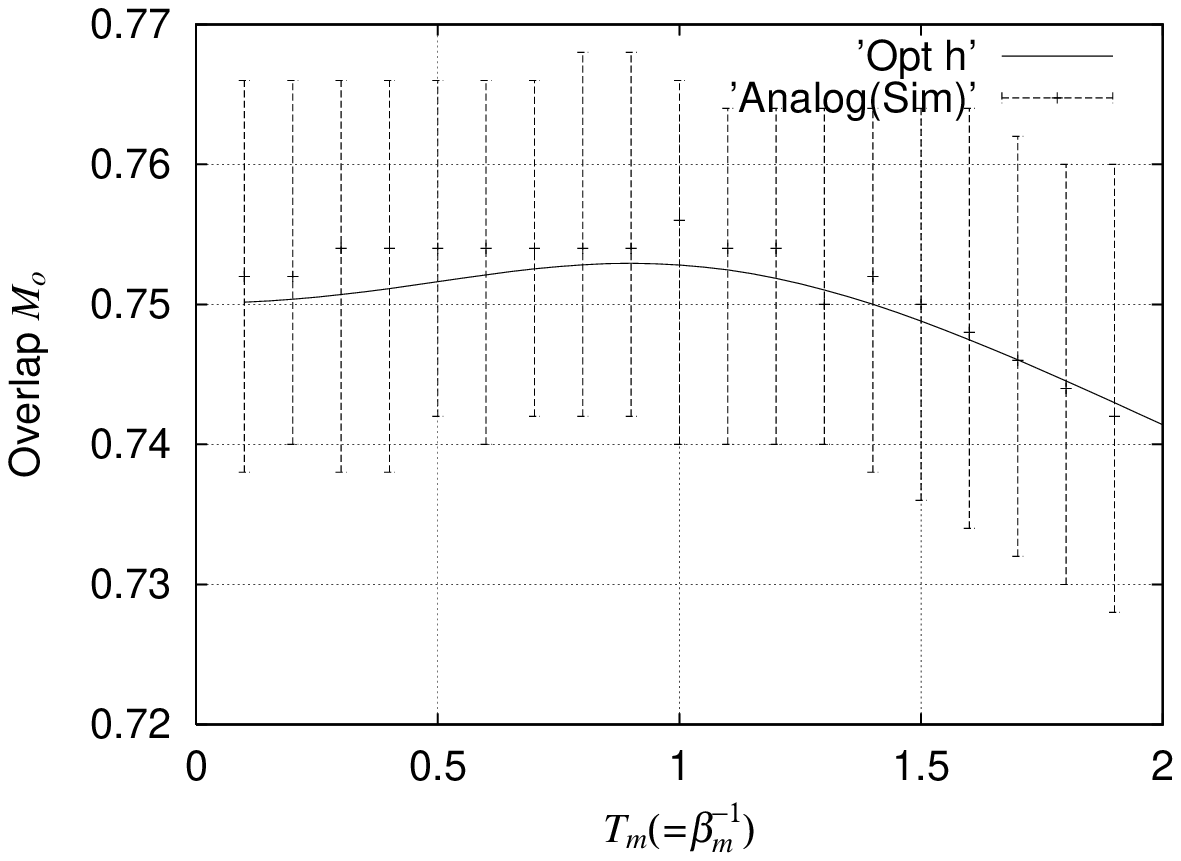}}
 \caption{
 Infinite range Ising model Restoration Ability without parity check:
 In decoding, when the proper temperature ($T_m = 0.9$) is chosen, 
 restoration ability indicates better than that of the temperature limit($T_m \rightarrow 0$).
 When decoding temperature $T_m$ equals to encoding temperature 
 $T_s = (\beta_s^{-1})$, which is appeared in the eq.
 (\ref{eq:orig_prior}), 
 the decoding ability becomes large.
 In the Monte Carlo simulation, 
 we use the infinite range Ising model, 
 which is described by eq. (\ref{eq:restoreprior}), as the decoding
 prior.
 \label{fig:isingrestore}
 }
\end{center}
\begin{center}
 \resizebox{!}{5.5cm}
 {\includegraphics{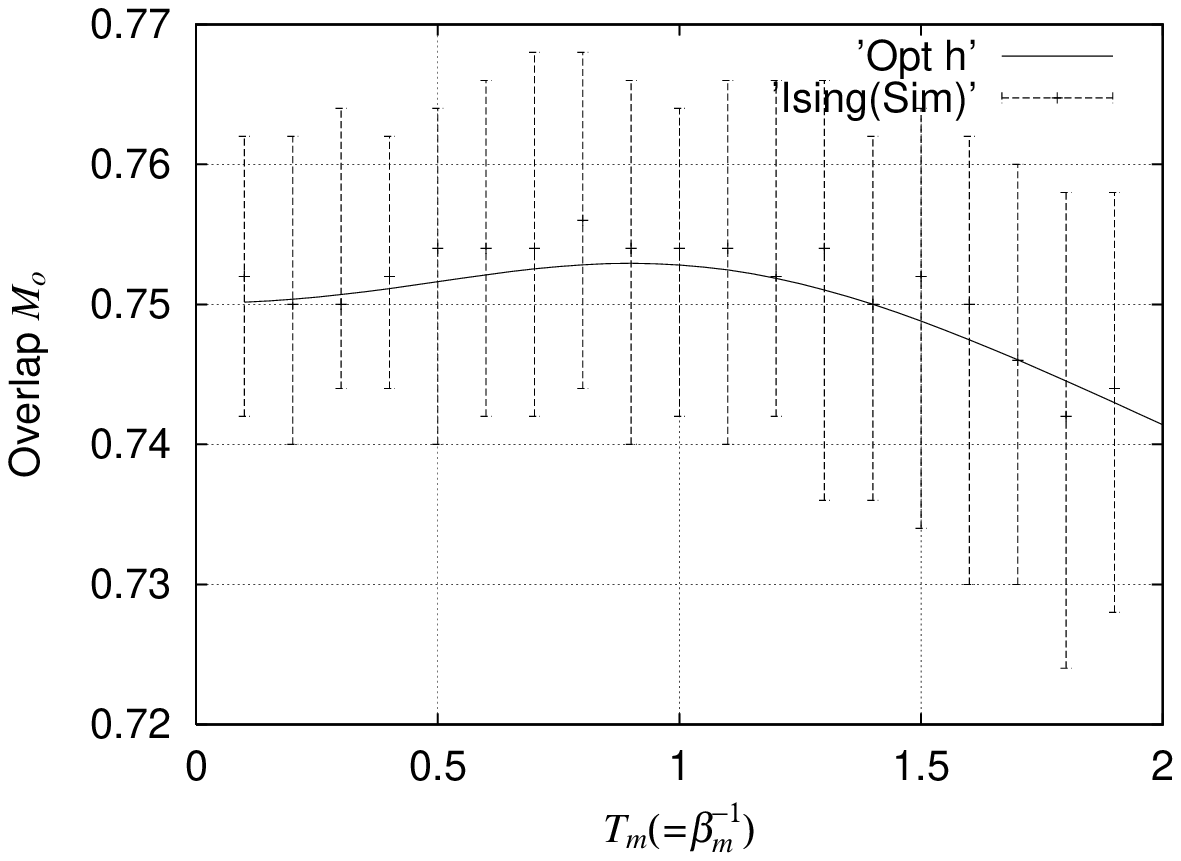}}
 \caption{
 Infinite range analog model Restoration Ability without parity check:
 Formally, the analysis result becomes identical to the infinite range 
 Ising model analysis result. 
 In the simulation, we use the eq. (\ref{eq:dynamics}) as the 
 discrete synchronous dynamics.
 }
 \label{fig:analogrestore}
\end{center}
\end{figure}

However, for image prior, 
the infinite range interaction is just a little strange assumption.
Thus, 
we introduced models whose prior have only nearest
neighbor interaction in \ref{sec:subsim4}.
We call them as Ising model with nearest neighbor interactions 
and 
analog model with nearest neighbor interactions respectively.
Unfortunately, it is difficult to treat the nearest neighbor interaction
models analytically.
Therefore, we calculated the result of nearest neighbor interaction models 
by computer simulation and compared them with the result of the infinite range models.
In the nearest neighbor interaction models, we assume that 
the corruption process in transmitting is same as infinite range model, 
that is defined by eq. (\ref{eq:noise}).

In the Ising model Monte Carlo simulation, 
we use the asynchronous Glauber dynamics, 
i.e. 
we selected one site $\sigma_i$ randomly, and calculate probability as
\begin{equation*}
 P( \sigma_i = \pm 1 ) =
  \frac{ (1 \pm 
  \tanh (\beta \sum_j J_{ij} \sigma_j + \frac{\beta_m}{N} \sum_j
  \sigma_j + h\tau_i) ) }{2}.
\end{equation*}
According to the probability the selected site $\sigma_i$ is decided.

On the other hand, for analog simulation,
we used discrete synchronous update rule 
described as eq. (\ref{eq:dynamics}).
i.e. whole units are updated simultaneously.
We consider 
each unit value in the equilibrium state 
as its thermal average.

\begin{figure}[t]   
\begin{center}
 \resizebox{!}{5.5cm}
 {\includegraphics{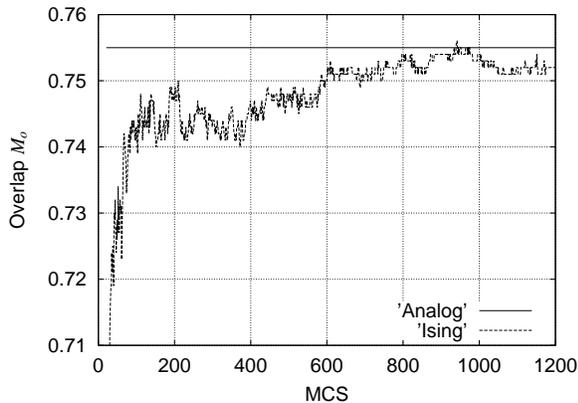}}
 \caption{Convergence time of Overlap: Ising vs Analog, 
 One MCS means $N_m = 2000$ updates, $T_s=T_m=0.9$, and $h=1$}
 \label{fig:speed}
\end{center}
\end{figure}

\begin{figure}[t]
\begin{center}
 \resizebox{!}{5.5cm}
 {\includegraphics{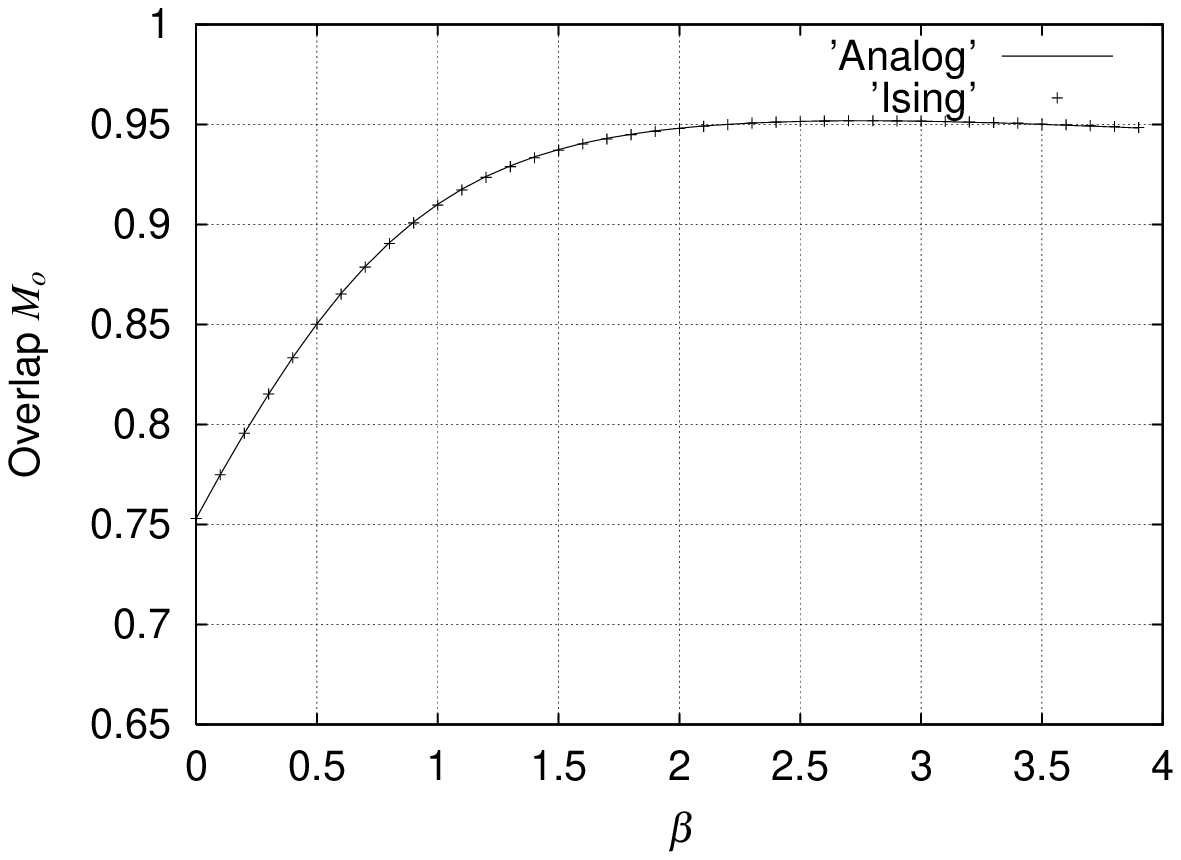}}
 \caption{
 Analysis comparison between infinite range analog model and infinite
 range Ising model using parity
 check code ($\beta > 0$) with small noise variance ($J=0.6$).
 For decoding, we assumed infinite range model that is described as
 eq. (\ref{eq:restoreprior}). In this figure, 
 the encoding and decoding temperatures is fixed as $T_s = T_m = 0.9$.
 }
 \label{fig:AvsI1}
\end{center}
\begin{center}
 \resizebox{!}{5.5cm}
 {\includegraphics{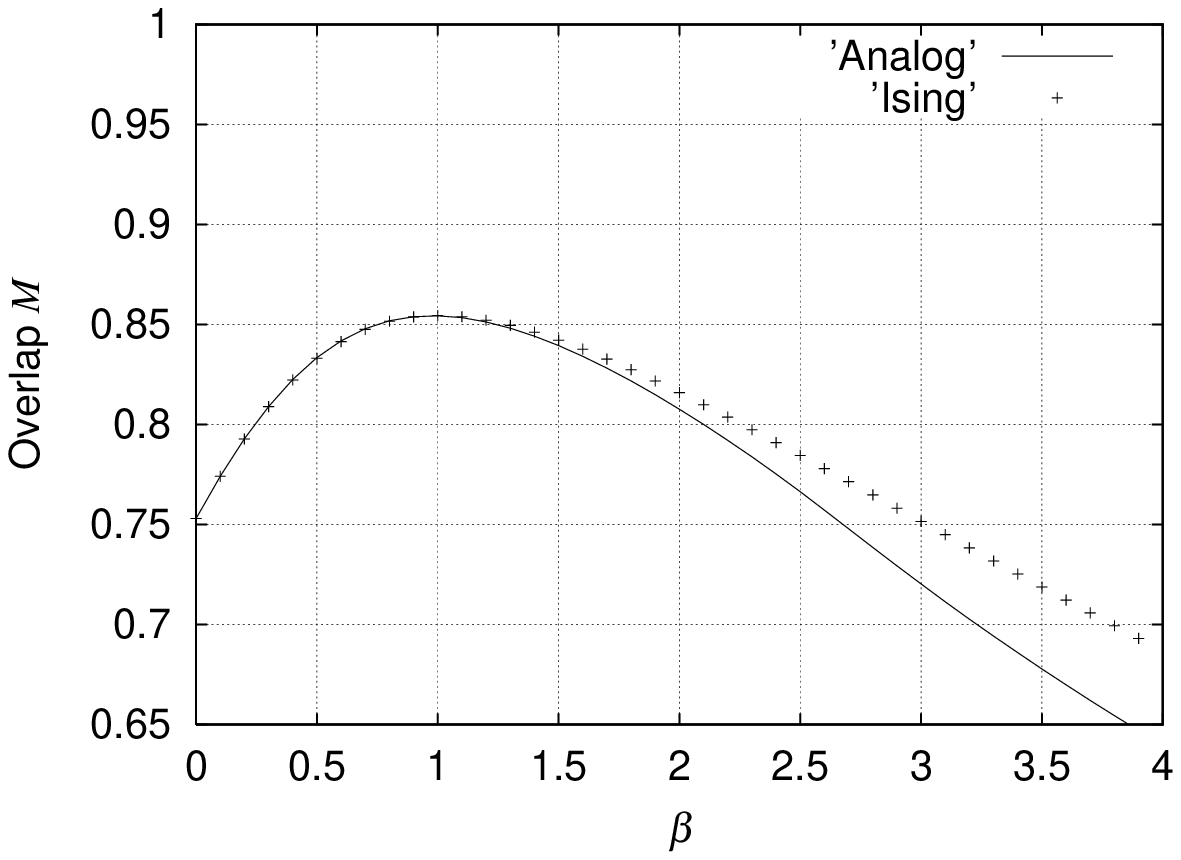}}
 \caption{
 Analysis comparison between infinite range analog model and infinite
 range Ising model using parity
 check code ($\beta > 0$) with large noise variance ($J=1.0$).
 For decoding, we assumed infinite range model that is described as
 eq. (\ref{eq:restoreprior}). In this figure, 
 the encoding and decoding temperatures is fixed as $T_s = T_m = 0.9$.
 }
 \label{fig:AvsI2}
\end{center}
\end{figure}
\begin{figure}[t]
\begin{center}
 \resizebox{!}{5.5cm}
 {\includegraphics{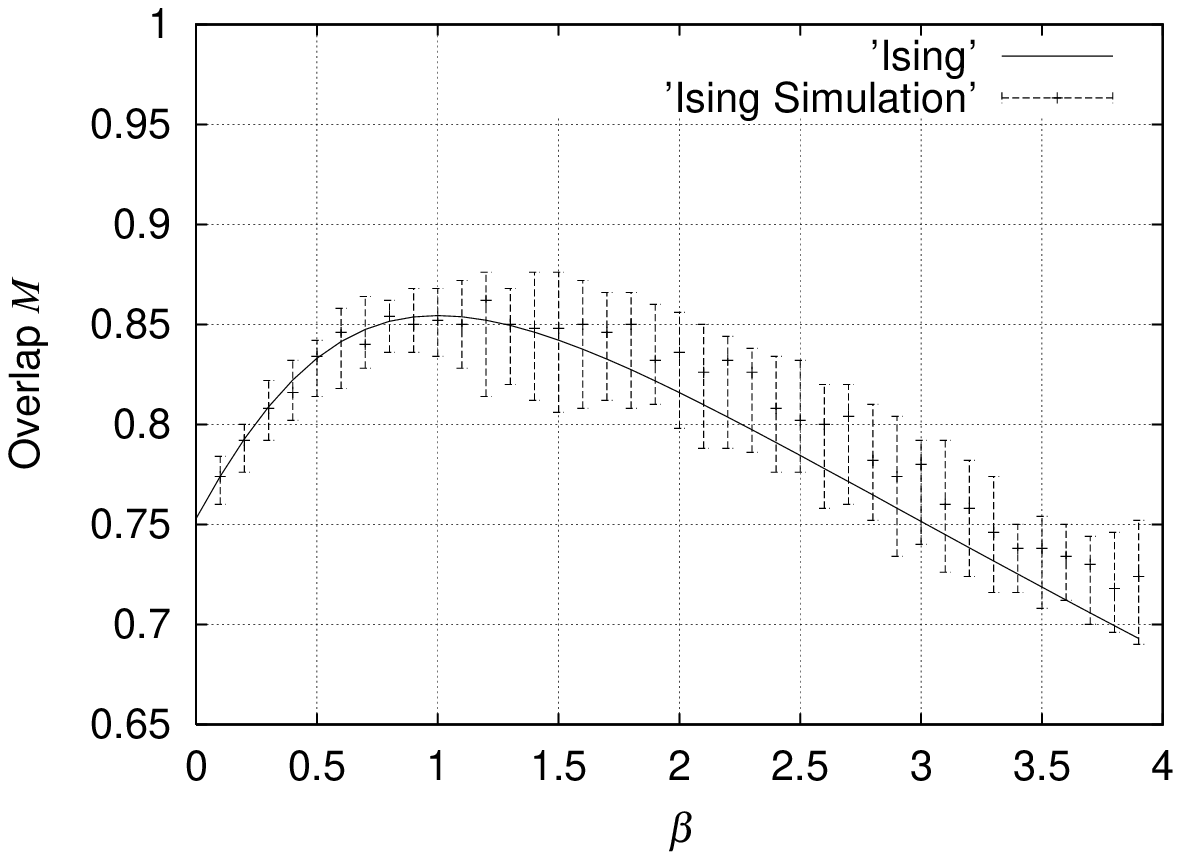}}
 \caption{Comparison between simulation result with infinite range
 Ising model restoration ability. 
 The parity check code is used and its noise level is defined as $J=1.0$.} 
 \label{fig:AvsI3}
\end{center}
\begin{center}
 \resizebox{!}{5.5cm}
 {\includegraphics{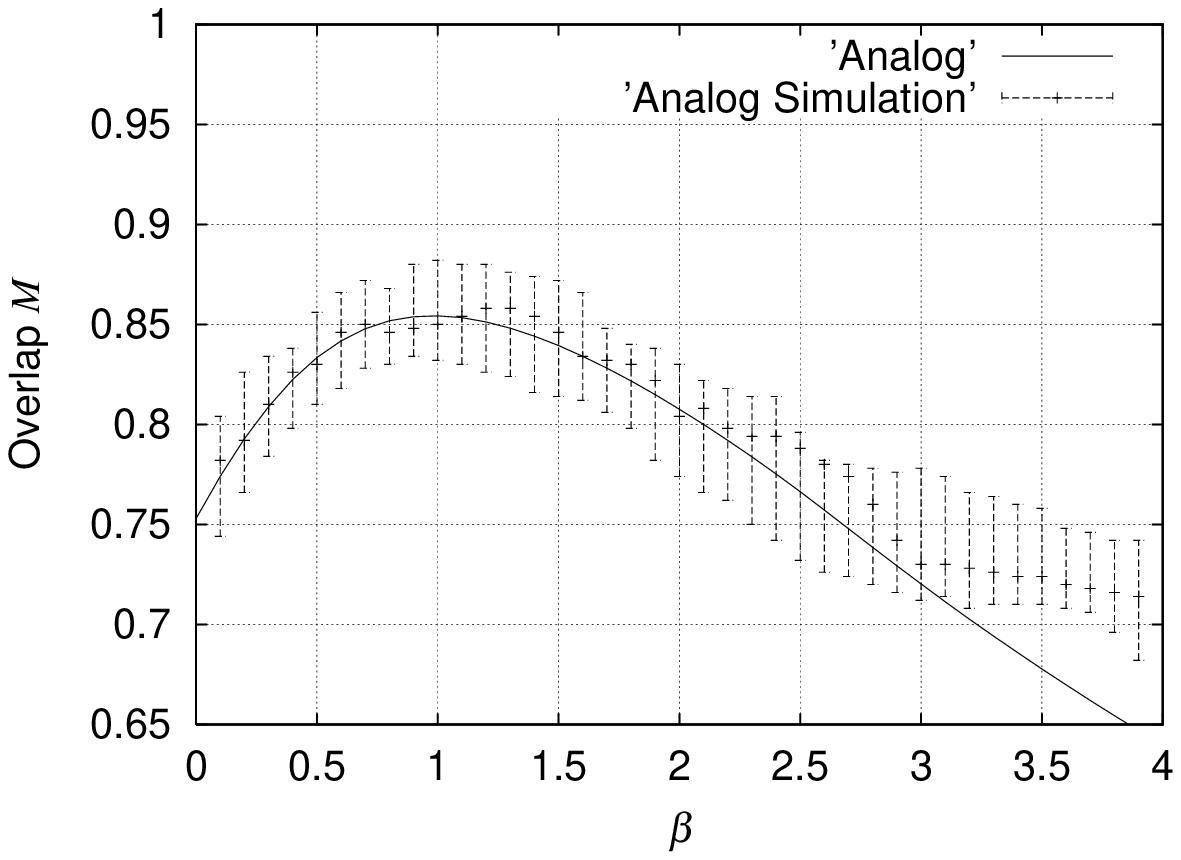}}
 \caption{Comparison between simulation result with the infinite range
 analog model restoration ability.
 The parity check code is also used and its noise level is defined as $J=1.0$.} 
 \label{fig:AvsI4}
\end{center}
\end{figure}

\subsection{Without parity check code}
\label{sec:subsim1}

First, we discuss the case where no parity check code is used
($\beta=0$).
This case corresponds to the conventional image restoration problem.
In the analysis, when $\beta = 0$, the order parameter equations
obtained in the previous section are identical to the result of 
the Ising analysis suggested by Nishimori \& Wong \cite{Nishimori00}.
Thus, the MPM inference by the stochastic infinite range Ising model and 
deterministic infinite range analog model are equivalent.
Figure \ref{fig:isingrestore} illustrates the restoring 
ability of finite temperature decoding using the infinite range Ising stochastic units.
Figure \ref{fig:analogrestore} shows the restoring ability using
infinite range analog model.
In each figure, the vertical axis means the overlap $M_o$, and
the horizontal axis means the restoring temperature $T_m$, which is 
corresponds to 
the restoring image prior coefficients $\beta_m$  
($T_m = \beta_m^{-1}$).
The solid line means the analysis result by the replica method, and each
error bar shows the quartile deviation obtained by the computer simulation.

We set the parameter $T_s$, which is the reciprocal of the original image
prior coefficients $\beta_s$ ($T_s = \beta_s^{-1}$), at $T_s = 0.9$.
At the limit of $T_m \rightarrow 0$, the MPM inference becomes
equivalent to the MAP inference.
In both figures \ref{fig:isingrestore} and \ref{fig:analogrestore}, 
the restoration ability around $T_m = (T_s) = 0.9$ is better 
than that of $T_m \rightarrow 0$.
This means that the MPM inference has
better image restoration ability 
rather than the MAP inference as far as the overlap.

\subsection{Convergence speed}
\label{sec:subsim2}

The biggest advantage of the analog model over the Ising model 
is its lower calculation cost.
In this section, we discuss the calculation time 
that should be set
in the computer simulation for each method.

We observed the convergence time of the overlap in each simulation.
Figure \ref{fig:speed} shows the convergence speed.
In the infinite range Ising model, we adopt the synchronous update and 
define $N_m$ updates as one Monte-Carlo step (1 MCS). 
$N_m$ means the number of units in the simulation.
A synchronous analog update corresponds to 1 MCS in the sense of
the number of update units.
The horizontal axis of Fig. \ref{fig:speed} indicates the
calculation time measured by MCS.
The vertical axis means the average of overlap at that time.
Each model achieved the macroscopic equilibrium state in 20 MCS.
The analog model adopts a deterministic algorithm, 
we can calculate $s^\infty_i$ at that time, 
and the $M_o$ converges at 20 MCS.
However, assuming an Ising stochastic algorithm, we should calculate the
thermal average of $s_i$ by sampling.
Figure \ref{fig:speed} shows that needs about 1000 MCS to converge.
In this result, 
the deterministic infinite range analog model converged 50 times faster than the
stochastic infinite range Ising model in finite temperature decoding.

\subsection{Using parity check code}
\label{sec:subsim3}

In the replica method analysis, the infinite range analog model and 
the infinite range Ising model are equivalent in the case using $\beta=0$.
This is derived from the fact that the susceptibility $\chi$ remained 
at $0$ under $\beta=0$.
Using the parity check code for better restoration, 
the analysis results differ.
In the case of $\beta>0$, $\chi$ becomes greater than $0$, and the term
corresponding to $\chi$ in eq. (\ref{eq:onsager}) is effective.
This term is called the Onsager reaction term in statistical mechanics.
As a result, a difference in analysis between the Ising model and the
infinite range analog model appears.

When the noise variance of the parity check code is small 
(e.g., $J\sim0.6$), 
the infinite range analog model and the infinite range Ising model are almost the same 
in the sense of overlap.
In this case, little noise exists in the parity check code, 
and better image restoration is possible.
Figure \ref{fig:AvsI1} shows a comparison of overlaps obtained by 
each replica analysis. 
The horizontal axis indicates the parameter $\beta$,
and the vertical axis indicates the overlap $M_o$.
We assumed that the decoding temperature is optimal ($T_m = T_s (= 0.9)$)
and that the parameter is $h=1$.
Comparing $\beta$ with $h$, $\beta$ becomes large, the restoring system tends
to rely on the parity check rather than the raw code. 
Each restoration result appears the same, and better restoration 
is achieved than in the case transmitting the raw code only.

Figure \ref{fig:AvsI2} shows the case of large noise variance in the parity
check code channel ($J=1.0$).
The horizontal axis signifies $\beta$.
In this case, each method restores better than when only the raw code
channel is used.
However, relying on $\beta$ too much results in poorer restoration result 
than when only the raw code channel is used.
A difference between the infinite range analog model and the infinite
range Ising model appears 
as restoration worsens.
When the noise variance of the parity check code becomes large, 
the restoration ability of infinite range analog model is not as good as 
that of the infinite range Ising model.

Figure \ref{fig:AvsI3} shows the computer simulation result of the
infinite range Ising model under the noise variance parameter $J=1.0$, and 
Fig. \ref{fig:AvsI4} shows the computer simulation result of the
infinite range analog model.
Each simulation matches the analysis curve, and the performance of 
infinite range analog model simulation is not much better than 
that of the infinite range Ising simulation.
However, when we properly set the ratio of the coefficients $\beta, \beta_m$, and $h$,
the performance of each is optimized and the peak of ability for each is
almost the same. 
Thus when we can estimate the hyper-parameters properly, 
there is no difference between 
these two methods.

\subsection{Ising/analog model with nearest neighbor interaction}
\label{sec:subsim4}

\begin{figure}[t]  
\begin{center}
 \resizebox{!}{5.5cm}
 {\includegraphics{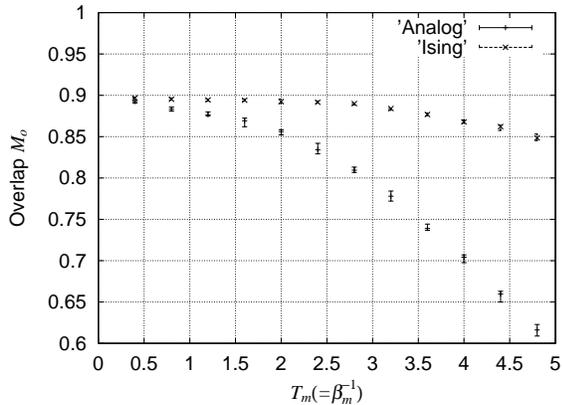}}
 \caption{Simulation result of Ising/analog model with nearest neighbor
 interactions.
 The vertical axis indicates overlap $M_o$, and horizontal axis means 
 decoding temperature $T_m = 1 / \beta_m$.
 }
 \label{fig:sim_2d_1}
\end{center}
\end{figure}

\begin{figure*}[t]   
\begin{center}
 \resizebox{0.95\textwidth}{!}
 {\includegraphics{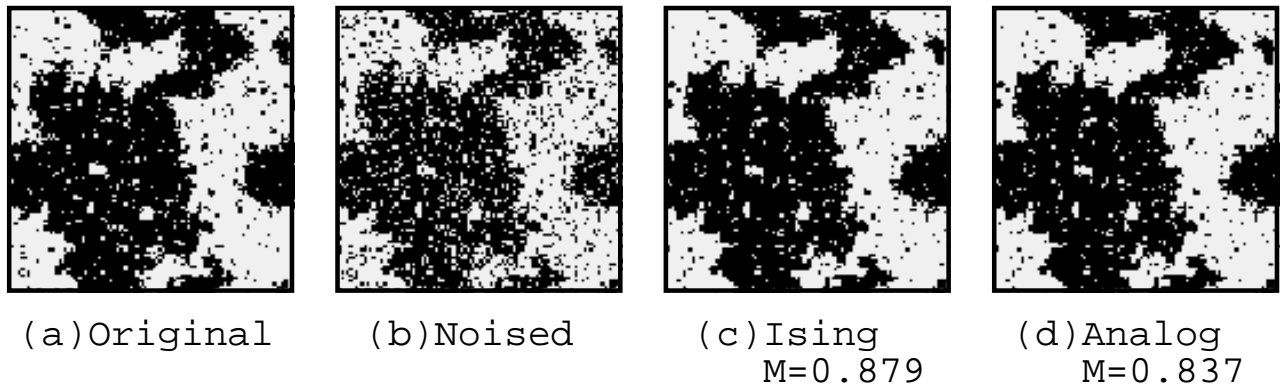}}
 \caption{Example of image restoration. The original image was
 created with a Ising/analog model with nearest neighbor prior model.
 (a) Original image; (b) Noised image; (c) Ising restoration result by 
 MPM estimation;
 and (d) Analog restoration result by MPM estimation}
 \label{fig:sim_2d_2}
\end{center}
\end{figure*}

In each infinite range model we compared, 
we assumed that all units have interactions with each other in prior
probabilities to use the mean field approximation.
However, it is more natural that these probabilities are defined by their
nearest-neighbor interactions. 
We call the model 
whose encoding/decoding prior probability is defined by the nearest neighbor interaction
on the pixel lattice 
the Ising/analog model with nearest neighbor interaction.
It is difficult to analyze the nearest neighbor interaction model with a statistical tool
because of its local interactions.
Nishimori \& Wong compared the analysis result of the Ising model  with nearest
neighbor interaction solved by computer simulation \cite{Nishimori00}.
Figure \ref{fig:sim_2d_1} shows the results of our reproduction.
The image size is $400 \times 400$ pixels, and 
the original image's prior probability is denoted:
\begin{eqnarray}
 P_s \left(\left\{ \xi_i \right\}\right) &=&
  \frac{1}{Z(\beta_s)}
  \exp( \beta_s \sum_{\langle ij \rangle} \xi_i \xi_j),
  \label{eq:2d_prior}
  \\
 Z(\beta_s) &=& \TR_{\mbox{\boldmath $\xi$}}
  \exp( \beta_s \sum_{\langle ij \rangle} \xi_i \xi_j),
\end{eqnarray}
where the summation  $\sum_{\langle ij \rangle}$ denotes a sum extending
over all pairs of neighboring sites on the pixel lattice.

In this simulation, we adopt $T_s=1/\beta_s=2.15$, 
generate the image $\{\xi_i\}$ by  the prior probability (\ref{eq:2d_prior}), 
and 
corrupt by flipping each pixel by the probability $p=0.1$.
We use only the raw code channel, not the parity check code channel
in this simulation.
By using the mean field theory, Nishimori \& Wong proved that
the overlap $M_o$ is maximized despite of the dimensions of the
system (image size), 
when the decoding temperature $T_m$ equals the image generating
temperature $T_s$.
In the simulation, we set $T_s = T_m = 2.15$.
Figure \ref{fig:sim_2d_1} shows the reproducing 
result of the Ising/analog model with nearest neighbor interaction
obtained by the computer simulation.
With the nearest-neighbor interaction model, it is clear that the overlap of the
analog model is smaller than that of the Ising model.
We believe that this phenomenon occurred because 
the analog model does not gather in the effect of fluctuation 
by other pixel values.

Thus, we should consider the nearest neighbor interaction prior effect 
in the computer simulation.
Figure \ref{fig:sim_2d_2} shows an example of the computer simulation
result.
Each image has an area of $400\times 400$ pixels.
Figure \ref{fig:sim_2d_2}(a) shows the original image.
In this image, noise-like patterns are inherently contained.
Figure \ref{fig:sim_2d_2}(b) shows the corrupted image.
Figure \ref{fig:sim_2d_2}(c) shows the restoration result by the Ising
model, and Figure \ref{fig:sim_2d_2}(d) shows the result by the analog
model.
The overlap of the Ising model (Figure \ref{fig:sim_2d_2}(c))
becomes $M_o = 0.879$, and the analog model (Figure \ref{fig:sim_2d_2}(c))
becomes $M_o = 0.837$.
In addition, when we applied the image restoration by MAP inference, 
the noise-like patterns contained in the original image
(Figure \ref{fig:sim_2d_2}(a)) were eliminated by the prior probability
term.
Thus, to maintain the fine structure of the original image, 
the MPM inference is an effective method.
This characteristics was confirmed in the analog model 
by the computer simulation.

\section{Discussion}
In this research, we discussed the image restoration problem
by finite temperature decoding.
In the MPM inference proposed by Nishimori \& Wong,
many trials are required  to calculate the thermal average for each unit.
We tried to replace this operation with
deterministic dynamics called naive mean field approximation.
At first, we applied mean field approximation to the infinite range
prior model
and analyzed this approximated system by the replica method.
We also quantitatively compared the restoration ability of the Ising
stochastic model with that of the analog deterministic model.
As a result,
we proved that their abilities are equivalent 
when we used only the raw code signal, 
that is, equivalent to conventional image restoration.
Assuming a redundant signal, called the parity check code,
the result of each method differed slightly.
However, when we adopted the ratio of the coefficients properly,
that is, the coefficients $\beta$, $h$, and $\beta_m$,
the restoring ability of each method showed no difference.
This means that if we can estimate these coefficients, called 
hyper-parameters, 
we can apply the analog deterministic method without 
worrying about the restoration ability.
Moreover, the analog deterministic method is $50$ times faster than the Ising
stochastic method. 

In the future problem, 
In analog model, 
Horiguchi pointed out that 
the inclination of output function around $\pm 1$
cause constitutive difference \cite{Horiguchi91}.
Thus, we should consider this effect.

Moreover, we will attempt to propose a more effective analog
approximation method for the nearest neighbor interaction model.

\end{document}